\documentclass[structabstract]{aa}  

\usepackage{amssymb,times,graphicx, subfigure}
\usepackage[english]{babel}
\usepackage[varg]{txfonts}
\usepackage{natbib,afterpage}
\bibpunct{(}{)}{;}{a}{,}{,}

\begin{document}
\title{PO and PN in the wind of the oxygen-rich AGB star IK~Tau\thanks{This work is partially based on observations carried out with the IRAM 30m Telescope. IRAM is supported by INSU/CNRS (France), MPG (Germany) and IGN (Spain).}}

   \author{E. De Beck\inst{1,2}
          \and T. Kami{\'n}ski\inst{1}
          \and N. A. Patel\inst{3}
          \and K. H. Young\inst{3}
          \and C. A. Gottlieb\inst{3}
          \and K. M. Menten\inst{1}
          \and L. Decin\inst{2,4}
          }

   \institute{Max-Planck-Institut f\"ur Radioastronomie, Auf dem H\"ugel 69, 53121 Bonn, Germany
                  \\ \email{edebeck@mpifr-bonn.mpg.de}
	      \and Instituut voor Sterrenkunde, Departement Natuurkunde en Sterrenkunde, Celestijnenlaan 200D, 3001 Heverlee, Belgium
	      \and Harvard-Smithsonian Center for Astrophysics, 60 Garden Street, MS78, Cambridge, MA 02138, USA
	      \and Sterrenkundig Instituut Anton Pannekoek, University of Amsterdam, Science Park 904, 1098 Amsterdam, The Netherlands 
             }

   \date{Received 22 February 2013 ; accepted 13 September 2013}

  \abstract
   {
   Phosphorus-bearing compounds have only been studied in the circumstellar environments of the asymptotic giant branch star IRC~+10\,216 and the protoplanetary nebula CRL~2688, both carbon-rich objects, and the oxygen-rich red supergiant VY~CMa. The current chemical models cannot reproduce the high abundances of PO and PN derived from observations of VY~CMa. No observations have been reported of phosphorus in the circumstellar envelopes of oxygen-rich asymptotic giant branch stars.
   }
   {
   We aim to set observational constraints on the phosphorous chemistry in the circumstellar envelopes of oxygen-rich asymptotic giant branch stars, by focussing on the Mira-type variable star IK~Tau.
   }
   {
   Using the IRAM~30\,m telescope and the Submillimeter Array, we observed four rotational transitions of PN ($J=2-1,3-2,6-5,7-6$) and four of PO ($J=5/2-3/2,7/2-5/2,13/2-11/2,15/2-13/2$). The IRAM~30\,m observations were dedicated line observations, while the Submillimeter Array data come from an unbiased spectral survey in the frequency range $279-355$\,GHz.
   }
   {
   We present the first detections of PN and PO in an oxygen-rich asymptotic giant branch star and estimate abundances    $X(\mathrm{PN/H_2})\approx3\times10^{-7}$ and $X(\mathrm{PO/H_2})$ in the range $0.5-6.0\times10^{-7}$. This is several orders of magnitude higher than what is found for the carbon-rich asymptotic giant branch star IRC~+10\,216. The diameter ($\lesssim$0\farcs7) of the PN and PO emission distributions measured in the interferometric data corresponds to a maximum radial extent of about 40 stellar radii. The abundances and the spatial occurrence of the molecules are in very good agreement with the results reported for VY~CMa. We did not detect PS or PH$_3$ in the survey.
   }
   {
   We suggest that PN and PO are the main carriers of phosphorus in the gas phase, with abundances possibly up to several $10^{-7}$. The current chemical models cannot account for this, underlining the strong need for updated chemical models that include phosphorous compounds.
   }


\maketitle
\titlerunning{PO and PN in the wind of the oxygen-rich AGB star IK~Tau}
\authorrunning{E. De Beck et al.} 


\section{Introduction}\label{sect:introduction}
Asymptotic giant branch (AGB) stars come in different chemical flavours ranging from oxygen-rich, over S-type, to carbon-rich stars. This sequence, which reflects progressing evolution, is characterised by an increasing abundance of atmospheric carbon (C), where the oxygen-rich chemistry (C/O$<$1) is assumed to correspond to the initial composition of AGB stars. 

From several surveys towards carbon-rich evolved stars, it has been established that the prevailing chemistry in their circumstellar envelopes (CSEs) is very rich, with molecules varying from the simple carbon monoxide (CO) to large cyanopolyynes and hydrocarbons, including polycyclic aromatic hydrocarbons \citep[PAHs;][]{boersma2006_C_pah,cherchneff2011_pah}. The carbon-rich AGB star most studied is without a doubt the nearby
\citep[$120-150\ \mathrm{pc}$; e.g.][]{loup1993,crosas1997,groenewegen2012_cwleo_distance,menten2012_ircdistance} and high mass-loss rate \citep[$1-4\times10^{-5}\ M_{\sun}\ \mathrm{yr}^{-1}$; e.g.][]{crosas1997,groenewegen1998,cernicharo2000_2mm,debeck2012_cch} object \object{IRC~+10\,216}, towards which many spectral line surveys have been executed in several different wavelength ranges \citep[e.g.][]{cernicharo2000_2mm,cernicharo2010_hifi1b,he2008,tenenbaum2010,patel2011_irc10216_sma}. 

The chemistry of oxygen-rich stars has been much less the subject of such surveys at high resolution and high sensitivity. Also, the CSEs of these stars seem to have a poorer chemical content than those of carbon-rich evolved stars \citep{olofsson2005_molabun}. This might, however, be largely due to an observational bias, caused by the lack of an O-rich star with a combination of mass-loss rate and distance comparable to those of IRC~+10\,216. Only high-sensitivity unbiased surveys of  several oxygen-rich stars will provide the much needed insight into their chemical content and the chemical evolution from oxygen-rich to carbon-rich in the latest phases of low and intermediate-mass stars. This paper focuses on the CSE of the oxygen-rich AGB star IK~Tau. This star loses mass at a rate of $\sim$$10^{-5}\ M_{\sun}\ \mathrm{yr}^{-1}$ in a wind that expands at a terminal velocity of $\sim$$18.5\ \mathrm{km\ s}^{-1}$ \citep{maercker2008,debeck2010_comdot,decin2010_iktau_nlte}. The CSE can be traced out to a distance of $\sim$27000\,AU from the central star, or slightly exceeding 100\arcsec\/ in radius, assuming the star is located at a distance of $250-265$\,pc \citep{decin2010_iktau_nlte}. The great similarity of its fundamental parameters (luminosity, effective temperature, mass-loss rate) to those of IRC~+10\,216 allows for a direct comparison of our results with those reported for the latter.

With a solar elemental abundance of $\mathrm{P/H}\approx3\times10^{-7}$ \citep{grevesse1998_solar}, phosphorus is an element only rarely observed in astrophysical environments. Thus far, HCP, PH$_3$, the CP and CCP radicals, PO, and PN are the only P-bearing molecules that have been identified around evolved stars. All of these, except PO, were observed towards the CSE of the carbon-rich AGB star IRC~+10\,216 \citep{agundez2007_hcp,agundez2008_ph3,guelin1990_irc10216_cp,halfen2008_ccp,milam2008_phosphorus}. HCP and PN were also observed in the envelope of the carbon-rich post-AGB object \object{CRL~2688} \citep{milam2008_phosphorus}, while only PO and PN have thus far been shown to be present in the CSE of an oxygen-rich evolved star, namely the red supergiant (RSG) VY~CMa \citep{milam2008_phosphorus,tenenbaum2007,ziurys2007_vycma_complexity,kaminski2013_vycma_sma}. 

We report on the PO and PN emission from the oxygen-rich AGB star IK~Tau detected in observations with the IRAM~30\,m telescope, and in an unbiased spectral line survey with the Submillimeter Array (SMA) in the spectral range $279-355$\,GHz. 


\begin{figure} 
\centering
\includegraphics[width=\linewidth]{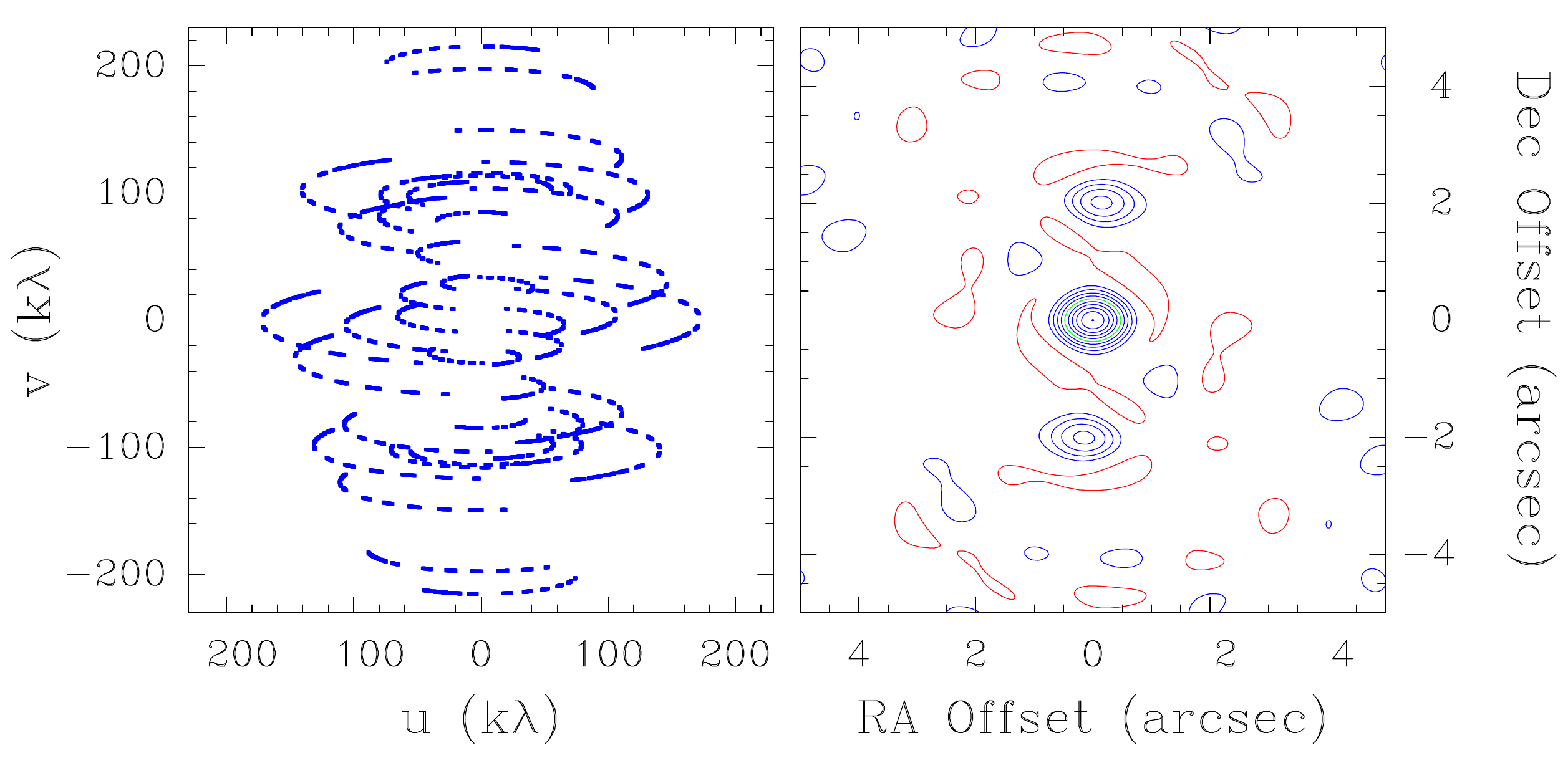}
\caption{Coverage of the $uv$-plane (\emph{left}) and shape of the synthesised beam (\emph{right}) in the SMA survey of IK~Tau, at 281.91\,GHz. The contours in the right panel are at 10\% spacings and start at the 10\% level. The 50\% contour is in green, all other positive contours are blue. The negative contours, marked in red, are at $-10\%$. }\label{fig:uvcoverage}
\end{figure}

\begin{figure}[htb] 
\centering
\includegraphics[width=.8\linewidth]{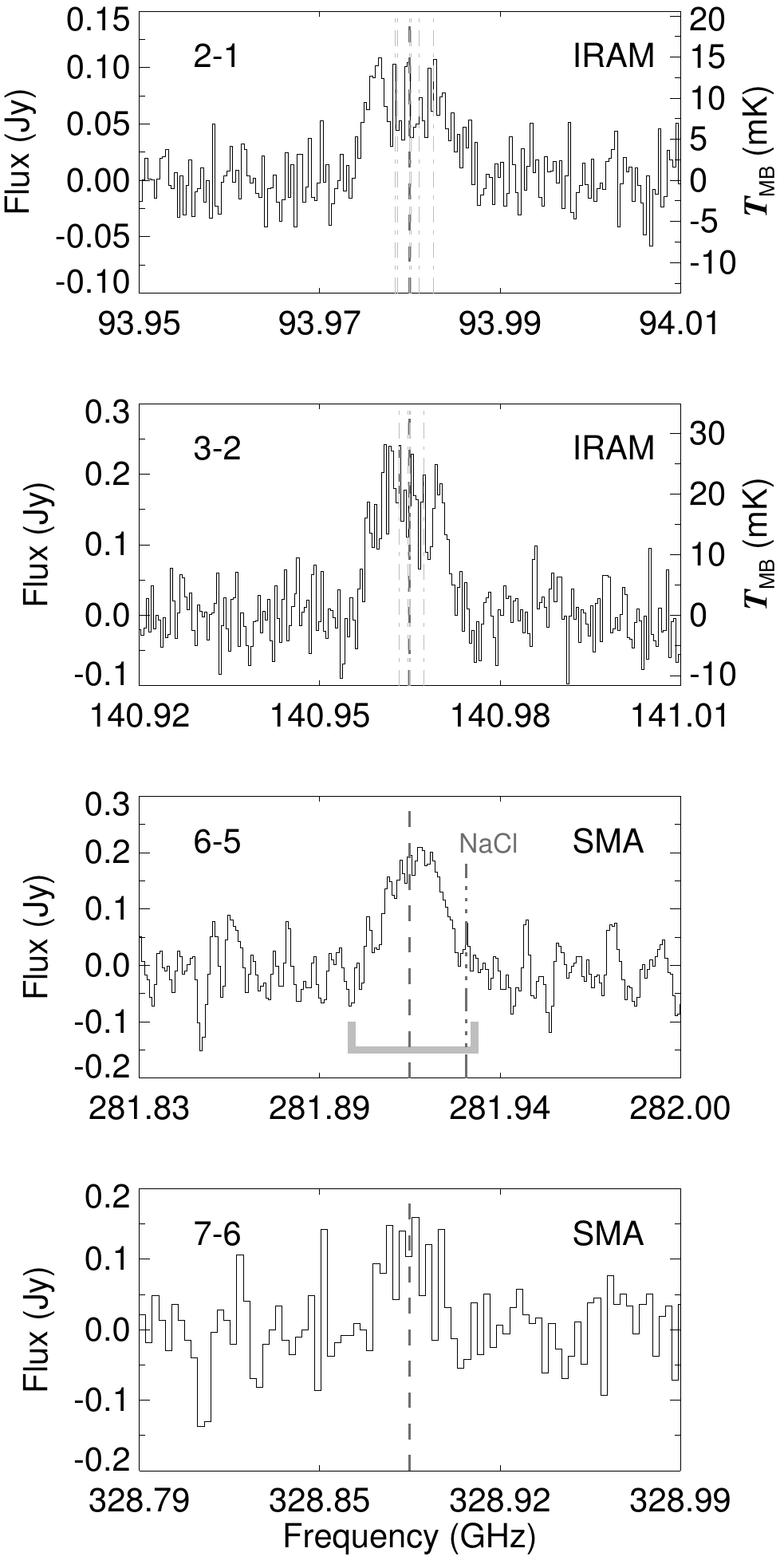}
\caption{PN emission from the CSE of IK~Tau. \emph{Upper two panels: } IRAM~30\,m data, \emph{lower two panels: }SMA survey data. The vertical dashed lines indicate the rest frequency of the respective rotational transition. For $J=2-1$ and $3-2$ we also indicate the position of the HFS components with dash-dotted lines. For $J=6-5$ we indicate the rest frequency of NaCl $\varv=2,J=22-21$, and the range over which was integrated to obtain the moment-zero map in Fig.~\ref{fig:PN65_momzero}.}
\label{fig:PN_SMA_IRAM_spectra}
\end{figure}

\begin{figure}[htb] 
\centering
\includegraphics[width=.9\linewidth]{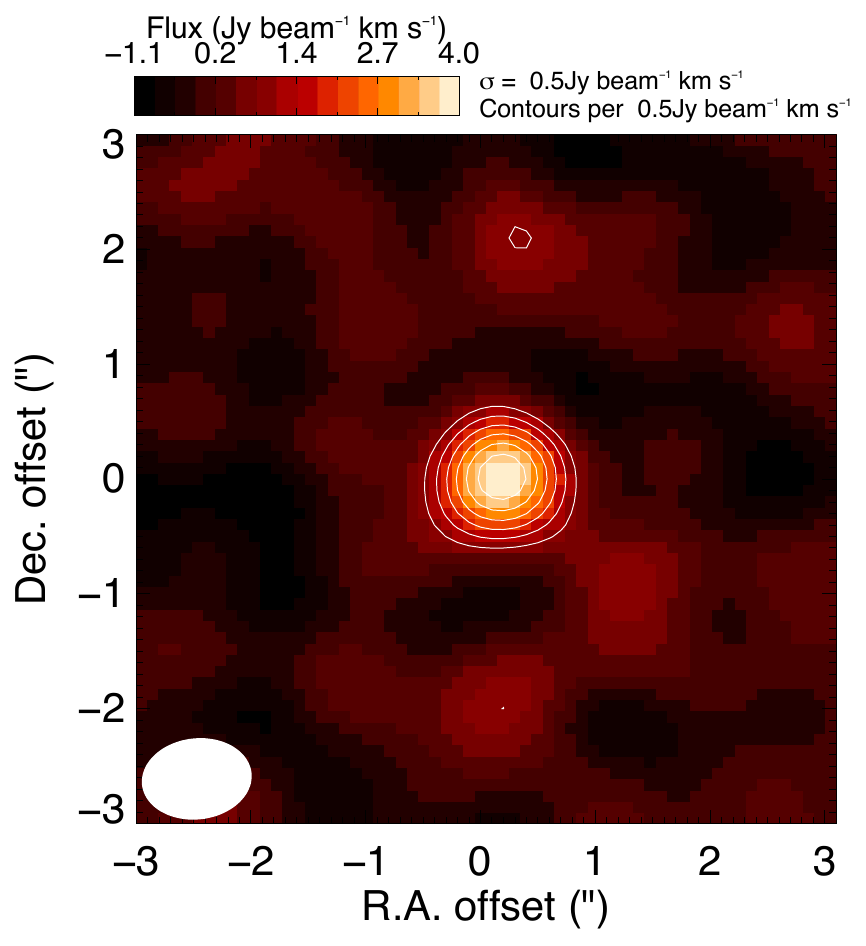}
\caption{Moment-zero map of the PN $J=6-5$ transition. The emission was integrated over the velocity range $[\varv_{\mathrm{LSR}}-22,\varv_{\mathrm{LSR}}+20]\ \mathrm{km\ s}^{-1}$. Contours start at $2\sigma$ ($1.0\ \mathrm{Jy \ beam}^{-1}\ \mathrm{km\ s}^{-1}$), with $1\sigma$ increments. The synthesised beam FWHM is represented by the filled ellipse in the bottom left corner.}\label{fig:PN65_momzero}
\end{figure}

\begin{figure} 
\includegraphics[width=\linewidth]{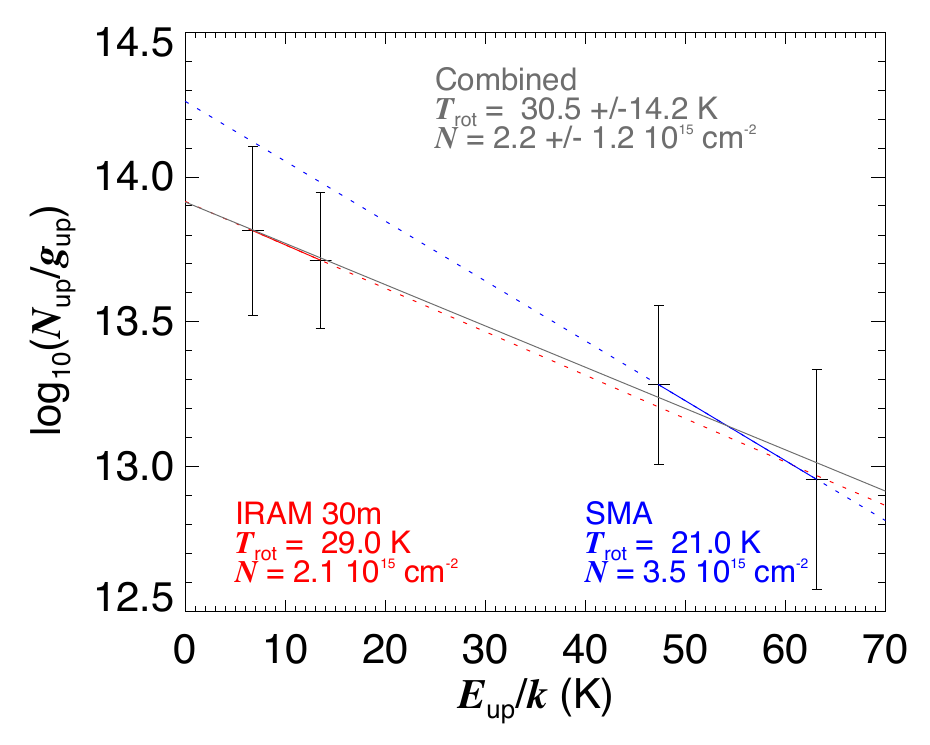}
\caption{Rotational temperature diagram for PN emission. The red and blue fits only take into account the IRAM~30\,m or the SMA data, respectively, while the grey fit is based on all four measurements.}\label{fig:PN_rotdiagram}
\end{figure}

\section{Observations}\label{sect:observations}
\subsection{IRAM~30\,m telescope: targeted observations}
Targeted line observations of PN $J=2-1,3-2$ and PO $J=5/2-3/2,7/2-5/2$  were performed on 2009 May 13, 14, and 15 using the EMIR heterodyne receiver system on the IRAM~30\,m telescope. The observations were carried out using wobbler switching with a 120\arcsec\/ beam throw, which resulted in flat spectral baselines.

The VESPA and WILMA autocorrelator backends were used, with spectral resolutions of $0.23-0.39$\,MHz and 2.0\,MHz, respectively. Due to the large spread in the frequency domain of the hyperfine structure present in the rotational transitions of PO (see Sect.~\ref{sect:po}), we limited our analysis of PO to the WILMA data. For the analysis of PN we used the high-resolution VESPA data.

Data processing was done using the GILDAS/CLASS package\footnote{http://www.iram.fr/IRAMFR/GILDAS}. Vertical and horizontal polarisation signals were averaged in order to reduce the noise. First-order polynomial baselines were subtracted from the spectra after the emission lines were masked, and the data were scaled by the main-beam efficiencies in order to obtain the main-beam temperature scale. The half-power beamwidth (HPBW) for the PN $J=2-1,3-2$ observations are 26\arcsec\/ and 17\arcsec, and 22\arcsec\/ and 16\arcsec\/ for PO $J=5/2-3/2,7/2-5/2$, respectively.

\subsection{SMA: unbiased survey covering 279-355\,GHz}
Between 2010 January 17 and February 2, we carried out a spectral survey in the range $279-355$\,GHz of IK~Tau using the SMA in the extended configuration. The spectral range of this survey covers two rotational transitions of PN ($J=6-5,7-6$) and two of PO ($J=13/2-11/2,15/2-13/2$). The tracks covering these transitions were taken on 2 February 2010 ($\sim$280\,GHz) and 22 January 2010 ($\sim$328\,GHz). Per observing track two windows of  4\,GHz separated by 8\,GHz were covered. The observations of IK~Tau were carried out in the same setup as and contemporaneously with the observations of VY~CMa. \cite{kaminski2013_tio_tio2} reported on some first results for the latter. Further results and more detailed descriptions of both surveys will be described in upcoming publications.

The shortest projected baseline in the extended configuration is 37\,m, the longest 225.8\,m, resulting in a characteristic spatial resolution at 345\,GHz of $\sim$0\farcs9 and to the observations not being sensitive to emission structures larger than about 5\arcsec. The $uv$-coverage of IK~Tau for the PN($J=6-5$) line at 281.9\,GHz is shown in the left panel of Fig.~\ref{fig:uvcoverage}. It is clear that, although the tracks were shared between IK~Tau and VY~CMa, and the hour-angle coverage was hence limited, the $uv$-coverage was still sufficient to result in a nice synthesised beam, as shown in the right panel of Fig.~\ref{fig:uvcoverage}.

The coordinates used for the phase tracking on IK~Tau are $\alpha_{\mathrm{J2000}}=03^{\mathrm{h}}53^{\mathrm{m}}28\fs866, \delta_{\mathrm{J2000}}=11^{\circ}24\arcmin22\farcs29$. Self-calibration could not be performed for the IK~Tau observations, because the central stellar continuum source was not strong enough for this purpose. Peak continuum fluxes in the survey vary around $0.05-0.10$\,Jy\,beam$^{-1}$ (per 2GHz window). Complex gain calibration for the IK~Tau observations was performed using observations of the quasar 0423-013. The spectral bandpass was calibrated using observations of 3C273 and flux calibration was obtained with observations of Vesta and Titan for the frequency ranges around 280\,GHz and 330\,GHz, respectively. For the 280\,GHz range the average system temperature, $T_{\mathrm{sys}}$, was 190\,K. For the 330\,GHz range $T_{\mathrm{sys}}$ averaged 140\,K, except for one antenna which had a $T_{\mathrm{sys}}$ of about 200\,K, on average. The correlator mode was configured with a uniform spectral resolution of 0.8\,MHz per channel.


\begin{table*} 
\caption{Observed PN lines.}\label{tbl:PN_stats}
\centering
\begin{tabular}{crcrrccccc}
\hline\hline\\[-2ex]
&&&&&&& Rms & Beamwidth or&Estimated\\
Transition &	\multicolumn{1}{c}{$\nu_{\mathrm{cat}}$}& \multicolumn{1}{c}{$E_{\mathrm{up}}/k$}	& \multicolumn{2}{c}{$\varv$-range}		& 	\multicolumn{1}{c}{$\delta\varv$}& \multicolumn{1}{c}{Line flux} & noise	 & aperture size&source size\\
$J^{\prime}-J$ & \multicolumn{1}{c}{(MHz)} & \multicolumn{1}{c}{(K)}&\multicolumn{2}{c}{($\mathrm{km\ s}^{-1}$)}	& ($\mathrm{km\ s}^{-1}$)	& \multicolumn{1}{c}{($\mathrm{Jy}\ \mathrm{km\ s}^{-1}$)} & \multicolumn{1}{c}{(mJy)} &\multicolumn{1}{c}{(\arcsec)} &\multicolumn{1}{c}{(\arcsec)}\\
\hline\\[-2ex]
\textbf{IRAM~30\,m}\\
$2-1$	&93979.8	& 6.8	& -21.97	& 18.73	& 1.00	&1.65	&20.1	&26&0.65\\
$3-2$	&140967.7	&  13.5  & -18.13	& 17.61	&0.83 	& 2.60	&25.4	&17&0.40\\
\textbf{SMA survey} \\
$6-5$	&281914.2	& 47.4	& -17.25	& 14.74	& 0.86 	& 4.04	&50.6	&$1\times1$&0.20\\
$7-6$	&328888.0	& 63.1	& -9.63		& 10.43	& 2.52	& 2.04	&52.5	&$1\times1$&0.15\\
\hline
\end{tabular}
\tablefoot{The respective columns list the observed transition, the rest frequency $\nu_{\mathrm{cat}}$ (obtained from CDMS), the upper-level energy $E_{\mathrm{up}}$, the velocity range covered by the emission features relative to the stellar LSR velocity, $\varv_{\mathrm{LSR}}$, i.e. $33.7\ \mathrm{km\ s}^{-1}$, the velocity resolution $\delta\varv$, the velocity-integrated intensity, i.e. total line flux, of the emission, the rms noise level at the given $\delta\varv$, the FWHM beamwidth or the aperture size corresponding to the (extracted) observations, and the size of the emission region estimated based on the velocity profile given by \cite{decin2010_iktau_nlte}.}
\end{table*}

\section{PN}\label{sect:pn}
Low-lying rotational transitions in the vibrational ground state of the closed-shell molecule PN ($X^1\Sigma^+$) show hyperfine structure (HFS) which is resolvable in laboratory measurements. This hyperfine splitting is caused by the interaction of the nuclear quadrupole moment of $^{14}$N with the electric field gradient at the nucleus \citep{raymonda1971,hoeft1972}. HFS in PN was neglected in the present study, because the component separation is much smaller than the intrinsic line widths in the source \citep{cazzoli2006}. Spectroscopic parameters such as frequencies and Einstein-$A$ coefficients were obtained from the standard entry in the Cologne Database for Molecular Spectroscopy \citep[CDMS;][]{mueller2001_cdms,mueller2005_cdms}.

We present the detection of PN emission in four rotational transitions. The $J=2-1$ and $3-2$ transitions were observed in the targeted line observations with IRAM~30\,m, while the $J=6-5$ and $7-6$ emission were measured in the SMA survey. Table~\ref{tbl:PN_stats} lists frequencies and energies above the ground state of these transitions, and the integrated fluxes and noise levels of the observations. Fig.~\ref{fig:PN_SMA_IRAM_spectra} shows all four observed transitions. We indicate the positions of the HFS components of the $J=2-1$ and $3-2$ transitions in Fig.~\ref{fig:PN_SMA_IRAM_spectra}. Several of the observed narrow peaks coincide very well with the catalogue frequencies of the HFS components. Interpreting the line shape is significantly complicated by this behaviour, and it is not unambiguous what type of profiles (parabolic, flat-topped, two-horned) these transitions exhibit. The IRAM~30\,m observations hence do not set straightforward constraints on the spatial extent or the optical thickness of the PN emission. The displayed SMA spectra for $J=6-5$ and $7-6$ were integrated over a square aperture of $1\farcs0\times1\farcs0$ in size, centred on the star. The $J=7-6$ emission is detected with a signal-to-noise ratio (S/N) of $\sim$3 at a velocity resolution of $5\ \mathrm{km\ s}^{-1}$, while the three lower-lying lines are detected with S/N in the range $5-11$ at velocity resolutions of $2.5-3.0\ \mathrm{km\ s}^{-1}$. Only in the case of PN $J=6-5$ is there potential contamination, owing to a narrow NaCl $\varv=2,J=22-21$ line at 281.932\,GHz. The total flux of this narrow line (full width at zero power FWZP\,$\approx3.5\ \mathrm{km\ s}^{-1}$), however, represents only $\sim$3\% of the integrated flux in the range $281.895-281.935$\,GHz. This was not further taken into account.

\subsection{Emitting region}\label{sect:pn_emittingregion}
We cannot directly constrain the size of the PN $J=2-1$ and $J=3-2$ emission regions from the IRAM~30\,m observations. The only available size estimate for the PN emitting region of an oxygen-rich evolved star is for VY~CMa (see Sect.~\ref{sect:comparison}). Based on this, we adopt an upper limit of $40\,R_{\star}$ for the radial extent of the $J=2-1$ emission region, corresponding to a 0\farcs65 source size. From the rotational-diagram analysis presented in the next section we conclude that this estimate is reasonable. Additionally, comparing the width of the $J=3-2$ line to the velocity profile of IK~Tau's wind presented by \cite{decin2010_iktau_nlte}, we derive a radial extent of $25\,R_{\star}$, equivalent to a 0\farcs40 source size for the gas emitting in this transition.

Fig.~\ref{fig:PN65_momzero} shows the integrated-intensity map (moment-zero map), integrated over the velocity range $[\varv_{\mathrm{LSR}}-22,\varv_{\mathrm{LSR}}+20]\ \mathrm{km\ s}^{-1}$, of the $J=6-5$ emission measured in the SMA survey.  Here $\varv_{\mathrm{LSR}}$ is IK Tauri's LSR velocity of $33.8\ \mathrm{km\ s}^{-1}$ \citep[e.g.][]{debeck2010_comdot}.

At the line's frequency, the restoring beam has a major axis full width at half maximum (FWHM) of 0\farcs960 and a minor axis FWHM of 0\farcs707, at a position angle of  $96.3^{\circ}$. From a 2-dimensional Gaussian fit to the moment-zero map, using the \texttt{JMFIT} task in the NRAO Astronomical Image Processing System (AIPS), we find that the emission distribution is unresolved. We derive an upper limit on the deconvolved source size of $0\farcs60\times0\farcs70$ with a position angle of $102^{\circ}\,^{+30^{\circ}}_{-24^{\circ}}$. This corresponds to a maximum diametrical extent of about $1.4\times10^{15}$\,cm\,$\approx87$ stellar radii ($R_{\star}$), assuming the stellar parameters listed by \citet[][see their Table~6]{debeck2010_comdot}. Comparing the detected velocity ranges to the velocity profile of \cite{decin2010_iktau_nlte}, we however derive much smaller upper limits to the emission-region sizes. For the $J=6-5$ and $J=7-6$ transitions we find, respectively, maximum radial extents of $\sim$$4\times10^{14}$\,cm ($\sim$$12\,R_{\star}$) and $\sim$$3\times10^{14}$\,cm ($\sim$$9\,R_{\star}$), i.e. maximum source sizes of 0\farcs20 and 0\farcs15.

\subsection{Rotational temperature diagram}
From the construction of a rotational temperature diagram \citep[][see their appendix]{snyder2005} based on the four emission lines presented in Fig.~\ref{fig:PN_SMA_IRAM_spectra} we can derive a rotational temperature of $30.5\pm14.2$\,K and a column density of $2.2\pm1.2\times10^{15}$\,cm$^{-2}$ for PN (see Fig.~\ref{fig:PN_rotdiagram}). This method assumes that the emission is optically thin and excited under conditions of local thermodynamic equilibrium (LTE).  We point out that the rotational diagram was constructed assuming different source sizes for the different transitions, in accord with the constraints presented in Sect.~\ref{sect:pn_emittingregion}. We note that the four observed lines cover only a small range in excitation energy ($\sim$$5-65$\,K) and as such leave large uncertainties on the derived values. Assuming a mass-loss rate of $1\times10^{-5}\ M_{\sun}\ \mathrm{yr}^{-1}$, a constant expansion velocity of $15\ \mathrm{km\ s}^{-1}$, and an abundance cut-off at $r_{\mathrm{max}}=1.25\times10^{15}$\,cm\,$\approx40\,R_{\star}$, we obtain an abundance $X(\mathrm{PN/H_2})=2.8\pm1.5\times10^{-7}$, suggesting that up to half of all available P could be locked into PN.

The assumption of optically thin emission might hold, but that of LTE probably does not. Kinetic temperature profiles for the CSE of IK~Tau, such as those presented for the radiative transfer models by e.g. \citet[][see their Fig. 4]{decin2010_iktau_nlte} and \citet[][see their Fig. 4]{maercker2008} give gas kinetic temperatures $T_{\mathrm{kin}}\gtrsim300$\,K for $r\lesssim 40\,R_{\star}$.   Based on the large difference we find here with the derived $T_{\mathrm{rot}}$, we suggest that PN is not excited under LTE conditions, although the rotational temperature diagram leads to an unambiguous fit. The PN lines are hence subthermally excited, since $T_{\mathrm{rot}}<<T_{\mathrm{kin}}$. The large dipole moment of PN \citep[$\mu_{\mathrm{PN}}=2.75$\,Debye; CDMS;][]{raymonda1971,hoeft1972} could explain this discrepancy between $T_{\mathrm{rot}}$ and $T_{\mathrm{kin}}$, in the sense that PN could be strongly affected by radiative excitation to its first vibrational state ($\varv=1$) at $\sim$7.5\,$\mu$m. To say more about this, however, observations of higher-$J$ transitions in the ground and excited vibrational levels are needed, and a detailed radiative transfer analysis of the PN emission needs to be carried out. This is outside the scope of this paper.

\begin{figure*}[htb] 
\sidecaption
\includegraphics[width=12cm]{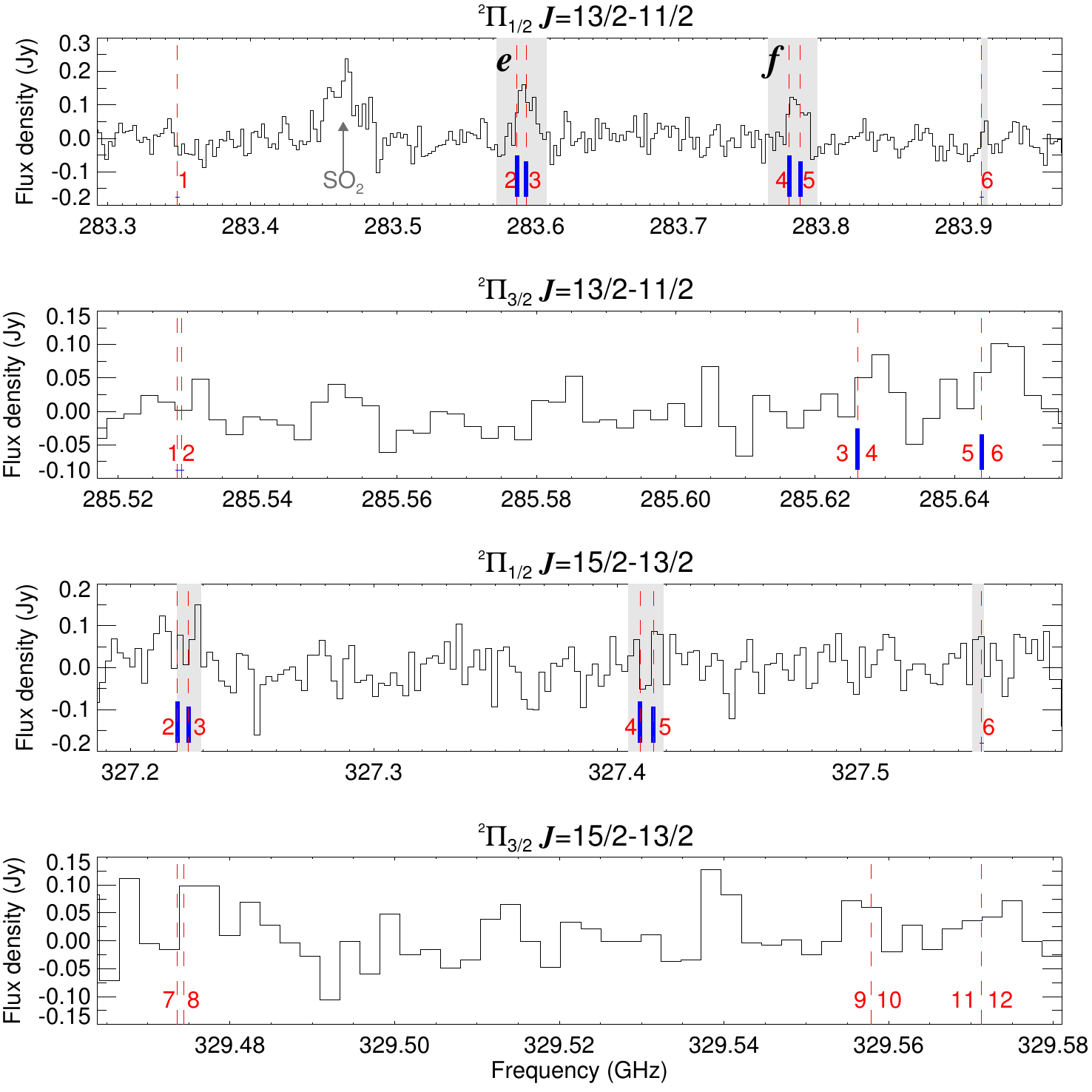}
\caption{Parts of the SMA survey spectrum corresponding to transitions of PO, ordered according to increasing frequency from top to bottom. All components are indicated as they appear in the last column of Table~\ref{tbl:PO_components}. Their theoretical intensities, relative to the strongest within the given spin component of the rotational transition, are indicated by the height of the blue bars. Note that the model predictions for some components are too weak to be plotted. The frequency ranges integrated to obtain the line fluxes used in the rotational diagram and listed in Table~\ref{tbl:PO_rotdiagram} are indicated with the grey-shaded areas. The $\Lambda$ components presented in Fig.~\ref{fig:PO7-6_momzero} are labelled with $e$ and $f$ in the top panel. Component 1 of the $^2\Pi_{1/2}\,J=15/2-13/2$ transition is not shown since it falls in a high-noise region of the spectrum. The top panel shows the contribution of the SO$_2$$(16_{0,16}-15_{1,15})$ transition at 283.465\,GHz.}
\label{fig:PO_components_SMA}
\end{figure*}

\begin{figure*}[htb] 
\centering
\subfigure[$\Lambda$ component $e$, centred at $\sim$283.59\,GHz.]{\includegraphics[width=.45\linewidth]{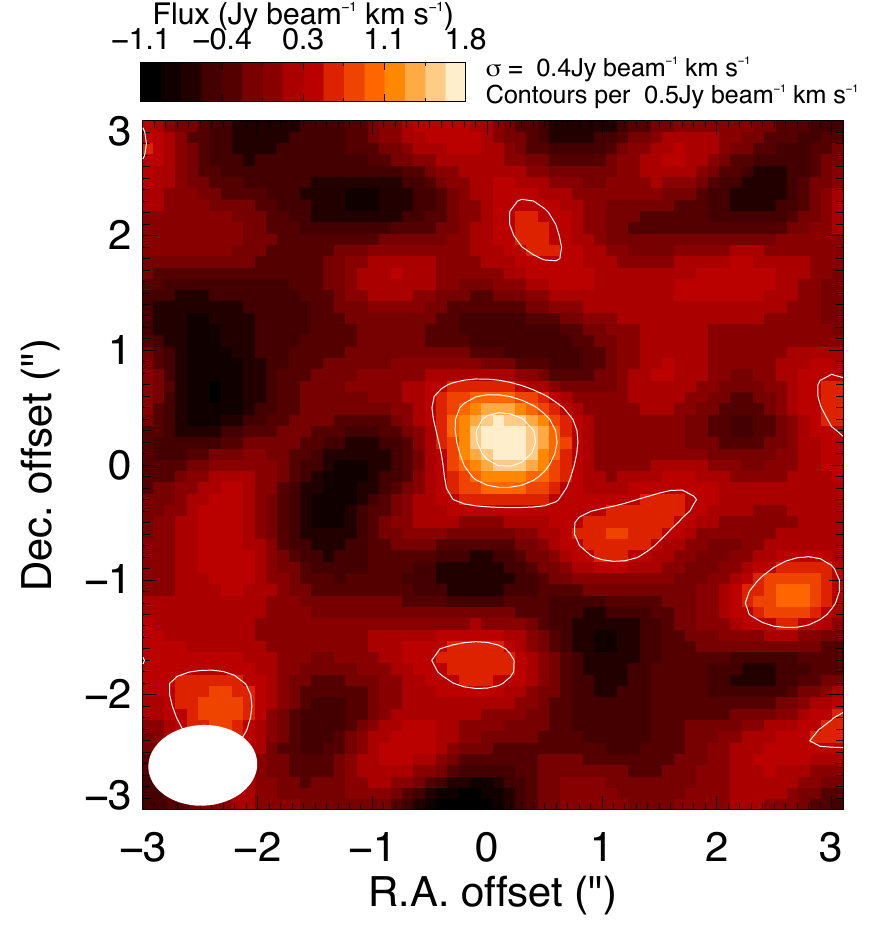}}
\subfigure[$\Lambda$ component $f$, centred at $\sim$283.78\,GHz.]{\includegraphics[width=.45\linewidth]{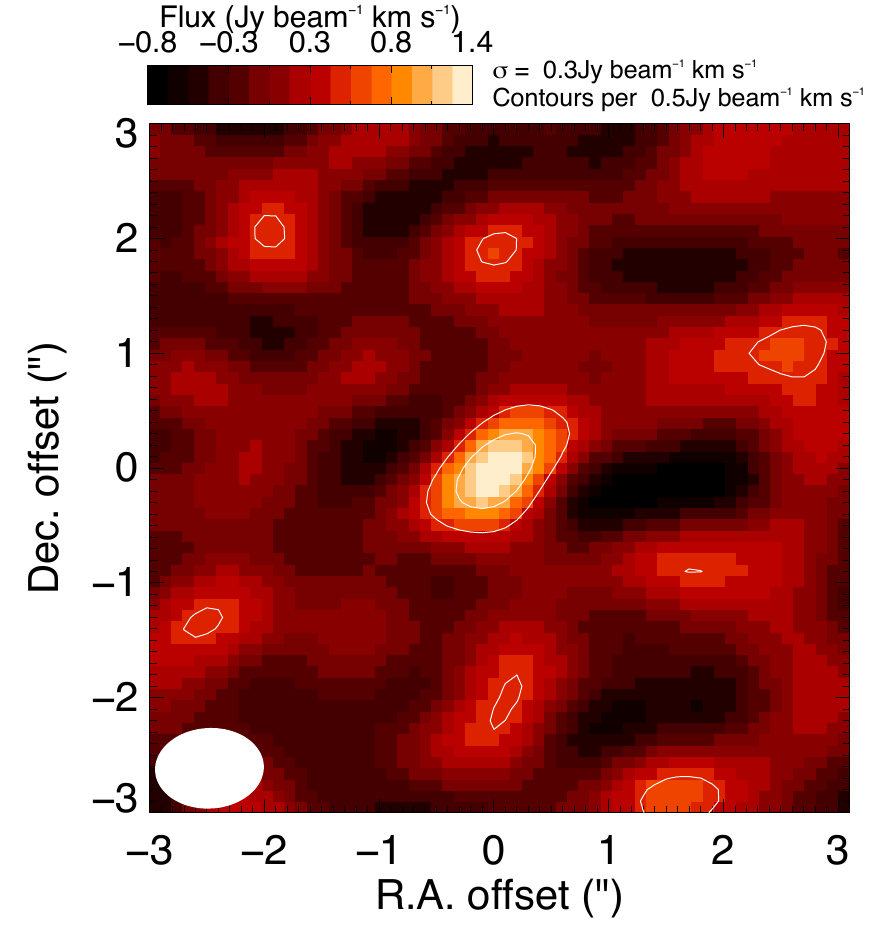}} 
\caption{Integrated intensity maps of the two strongest PO ($^2\Pi_{1/2} \,J = 13/2 - 11/2$) emission features.  Contours start at $0.5\ \mathrm{Jy\ beam}^{-1}\ \mathrm{km\ s}^{-1}$ ($1.7\sigma$), with $0.5\ \mathrm{Jy\ beam}^{-1}\ \mathrm{km\ s}^{-1}$ increments. The synthesised beam FWHM is represented by the filled ellipse in the bottom left corner. }\label{fig:PO7-6_momzero}
\end{figure*}

\subsection{Comparison to IRC~+10\,216 and VY~CMa}\label{sect:comparison}
\cite{patel2011_irc10216_sma} identified emission from the PN $J=7-6$ transition in the SMA survey of IRC~+10\,216 as line 251, while the PN $J=6-5$ line was not covered in their survey ($294-355$\,GHz). The line flux in the velocity range $[\varv_{\mathrm{LSR}}-20,\varv_{\mathrm{LSR}}+20]\ \mathrm{km\ s}^{-1}$ is $\sim$5\,Jy. Assuming a distance of 150\,pc and a mass-loss rate of $2\times10^{-5}\ M_{\sun}\ \mathrm{yr}^{-1}$ for IRC~+10\,216, and a distance of 260\,pc and a mass-loss rate of $1\times10^{-5}\ M_{\sun}\ \mathrm{yr}^{-1}$ for IK~Tau, we find that the IK~Tau $J=7-6$ emission (Table~\ref{tbl:PN_stats}) is approximately 2.5 times as strong.

Based on the analysis of three lines (one of which was blended with a $^{30}$SiC$_2$ emission line) \cite{milam2008_phosphorus} derived a column density of $1\times10^{13}$\,cm$^{-2}$ for IRC~+10\,216, corresponding to an abundance $X(\mathrm{PN}/\mathrm{H}_2)\approx3\times10^{-10}$ for a mass-loss rate of $3\times10^{-5}\ M_{\sun}\ \mathrm{yr}^{-1}$. \cite{agundez2007_hcp} found a column density of $1.6\times10^{12}$\,cm$^{-2}$, i.e. almost an order of magnitude lower. \cite{milam2008_phosphorus} derived an $e$-folding radius of $4\times10^{16}$\,cm ($\approx$36\arcsec) for a Gaussian PN abundance profile, while \cite{agundez2007_hcp} derived a distribution of PN peaking in a shell at $\sim$$20-30$\arcsec\/ or $\sim$$1000-1500\,R_{\star}$ (see their Fig.~3). \cite{agundez2007_hcp} proposed that the formation of PN in the CSE of IRC~+10\,216 follows from neutral-neutral reactions in the outer wind. These reactions involve CP and CN -- photodissociation products of HCP and HCN -- and are hence probably not as significant in an O-rich CSE. Moreover, the SMA observations of IK~Tau support the formation of PN in the inner regions of the CSE, at a few $R_{\star}$, as opposed to what was derived for IRC~+10\,216.

Combining the single-dish data of PN in VY~CMa of \cite{milam2008_phosphorus} with the SMA data obtained in the survey mentioned earlier \citep[][submitted]{kaminski2013_vycma_sma} and from \cite{fu2012}, we derive a column density of $3\times10^{15}$\,cm$^{-2}$. Assuming a mass-loss rate of $10^{-4}\ M_{\sun}\ \mathrm{yr}^{-1}$ and a constant expansion velocity of $35\ \mathrm{km\ s}^{-1}$, this leads to an abundance $X(\mathrm{PN/H_2})\approx1\times10^{-8}$. The PN emitting region in the case of VY~CMa covers a radial extent out to $\sim$$41\,R_{\star}$.

Our derived abundance $X(\mathrm{PN/H_2})\approx3\times10^{-7}$ for IK~Tau and the newly derived $X(\mathrm{PN}/\mathrm{H}_2)\approx1\times10^{-8}$ for VY~CMa imply that PN is by roughly two to three orders of magnitude more abundant in O-rich environments than in C-rich environments. Since there is no reason to assume a different atmospheric abundance of P, and given the similar stellar and wind parameters of IK~Tau and IRC~+10\,216, we conclude that there must be a fundamental difference in the phosphorous chemistry between the respective chemical types, which influences both the abundance and the distribution of PN in the CSE. In Sect.~\ref{sect:discussion} we further discuss the observed and predicted abundances of different P-bearing species.


\begin{figure*}[htb] 
\sidecaption
\includegraphics[width=12cm]{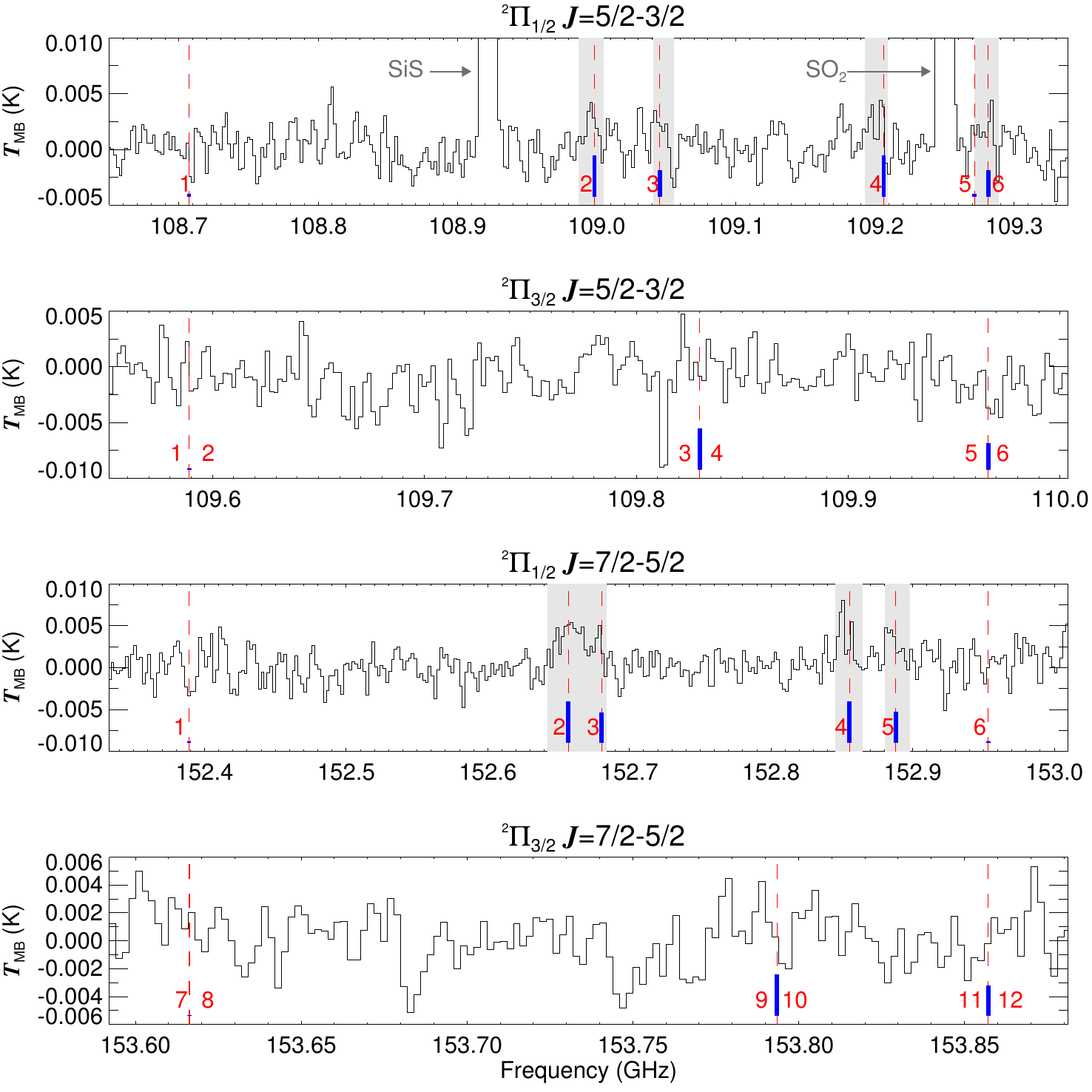}
\caption{IRAM~30\,m observations of PO, ordered according to increasing frequency from top to bottom. All components are indicated as they appear in the last column of Table~\ref{tbl:PO_components}. Their theoretical intensities, relative to the strongest within the given spin component of the rotational transition, are indicated by the height of the blue bars. Note that the model predictions for some components are too weak to be plotted. The frequency ranges integrated to obtain the line fluxes used in the rotational diagram and listed in Table~\ref{tbl:PO_rotdiagram} are indicated with the grey-shaded areas.}
\label{fig:PO_components_IRAM}
\end{figure*}

\section{PO}\label{sect:po}
The rotational spectrum of the vibrational ground state of PO ($X^2\Pi_r$) in the submillimetre range is characterised by fine and hyperfine structure. The lower and upper spin components, $^2\Pi_{1/2}$ and $^2\Pi_{3/2}$, respectively, of every rotational transition $J^{\prime}-J$ are separated by $\sim$320\,K in energy. The HFS components correspond to the transitions $F^{\prime}-F$ that belong to either the $e$ or $f$ $\Lambda$ component. The intrinsically strongest components for any rotational transition are those with $F^{\prime}-F=1$ in the $^2\Pi_{1/2}$ component. See Table~\ref{tbl:PO_components} for the rest frequencies, excitation energies and Einstein-$A$ coefficients of the observed transitions \cite[as taken from CDMS, following][]{kawaguchi1983_po,kanata1988_po_dipolemoment,bailleux2002_po}.

PO was for the first time detected in space by \cite{tenenbaum2007} who measured the $\Lambda$ components for  $^2\Pi_{1/2}\,J=11/2-9/2$ and $^2\Pi_{1/2}\,J=13/2-11/2$ at $\sim$240\,GHz and $\sim$283.5\,GHz, respectively, towards VY~CMa. We now present the first detections of PO towards an O-rich AGB star. 

Fig.~\ref{fig:PO_components_SMA} shows the PO $^2\Pi_{1/2}\,J=13/2-11/2$ and $^2\Pi_{1/2}\,J=15/2-13/2$ spectra of IK~Tau measured with the SMA, extracted for a $1\arcsec\times1\arcsec$ area around the central star.  The $F^{\prime}-F=1$ components of $^2\Pi_{1/2}\,J=13/2-11/2$ with an excitation energy of $\sim$50\,K are clearly detected and moment-zero maps of these components are shown in Fig.~\ref{fig:PO7-6_momzero}.  In these maps, both components are detected with a $\mathrm{S/N}\approx4.5$ and we find that the emission is completely unresolved by the SMA beam. The detection of the $^2\Pi_{1/2}\,J=15/2-13/2$ components at $\sim$66\,K is uncertain owing to high noise at this frequency. The higher-energy $^2\Pi_{3/2}\,J=13/2-11/2$ and $^2\Pi_{3/2}\,J=15/2-13/2$ transitions, at 370.8\,K and 386.7\,K, respectively, were not detected.

The IRAM~30\,m observations of the $^2\Pi_{1/2}\,J=5/2-3/2,7/2-5/2$ transitions (see Fig.~\ref{fig:PO_components_IRAM}) bear further evidence for the presence of PO in the CSE of IK~Tau. Analogous to what was seen for the SMA observations, the lower-energy $^2\Pi_{1/2}$ components are detected, while the $^2\Pi_{3/2}$ components are not detected with the present sensitivity.

Owing to the low S/N we cannot firmly establish the velocity range spanned by the PO emission lines in the SMA data, but we find velocities  not reaching more than $\approx$$7\ \mathrm{km\ s}^{-1}$, implying that the observed emission most likely does not arise from the fully accelerated wind. The IRAM~30\,m spectra, on the other hand, indicate that the emission could arise from a part of the wind that is fully accelerated. However, the interpretation of the data is not unambiguous, owing to the blends of the HFS components, the rather low S/N values, and the low velocity resolution ($3.9-5.5\ \mathrm{km\ s}^{-1}$).

Based on the interferometric maps and the comparison of the measured velocities with the velocity profile for IK~Tau's wind presented by \cite{decin2010_iktau_nlte}, we derive a maximum radial extent of the PO emission distribution of $1.25\times10^{15}$\,cm\,$\approx40\,R_{\star}$.

\subsection{Rotational temperature diagram}\label{sect:po_rotdiagram}
\begin{figure}[htb] 
\includegraphics[width=\linewidth]{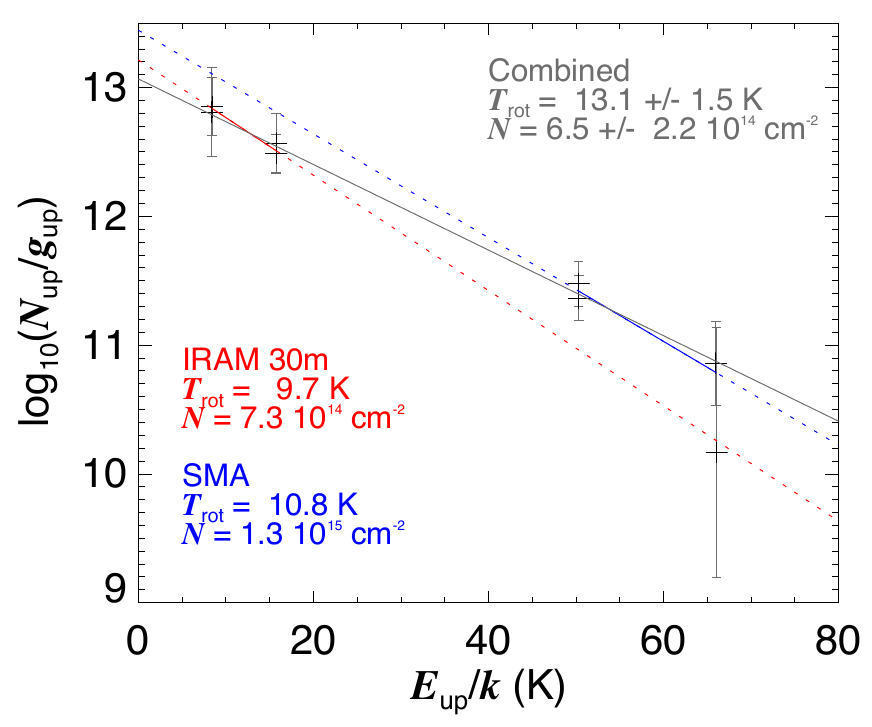}
\caption{Rotational temperature diagram for PO  $^2\Pi_{1/2}$ emission measured in the IRAM~30\,m and SMA data. The error bars represent $1\sigma$ uncertainties in the data. The red and blue fits only take into account the IRAM~30\,m or the SMA data, respectively, while the grey fit combines these.}
\label{fig:PO_rotdiagram}
\end{figure}

In Fig.~\ref{fig:PO_rotdiagram} we show a rotational temperature diagram for the PO $^2\Pi_{1/2}$ emission measured in the IRAM~30\,m and SMA data. For all transitions we assumed the same spatial extent of the emitting region. For every $^2\Pi_{1/2}\,J$ transition, we treat the two $\Lambda$ components $e$ and $f$ separately.   When blends are involved, we split the line intensity over the different HFS components according to their theoretical line strengths. Since the first HFS component of $^2\Pi_{1/2}\,J=15/2-13/2$ falls in the spectral window heavily contaminated by the atmospheric H$_2$O absorption line at $\sim$325\,GHz, we assume that it has the same integrated intensity as component number 6 (at 327.550\,GHz) of this rotational transition. These HFS components, belonging to the $f$ and $e$ $\Lambda$ components, respectively, have identical theoretical line strengths. We also assume the same noise characteristics. Relevant line fluxes, noise values, and frequency resolutions are stated in Table~\ref{tbl:PO_rotdiagram}. The ranges integrated to obtain these line fluxes are indicated in Figs.~\ref{fig:PO_components_SMA} and \ref{fig:PO_components_IRAM}. Considering all of the above, we wish to be cautious in interpreting the obtained results. For the IRAM~30\,m data we find $T_{\mathrm{rot}}=9.7\pm6.7$\,K and $N_{\mathrm{col}}=7.3\pm7.3\times10^{14}$\,cm$^{-2}$, while the SMA data lead to $T_{\mathrm{rot}}=10.8\pm5.7$\,K and $N_{\mathrm{col}}=1.3\pm3.5\times10^{15}$\,cm$^{-2}$. Combining the two data sets leads to $T_{\mathrm{rot}}=13.1\pm1.5$\,K and $N_{\mathrm{col}}=6.5\pm2.2\times10^{14}$\,cm$^{-2}$. Omission of the $J=15/2-13/2$ data points leads to an insignificant change of the rotational temperature and column density in the combined fit. Assuming a mass-loss rate of $1\times10^{-5}\ M_{\sun}\ \mathrm{yr}^{-1}$, a constant expansion velocity of $15\ \mathrm{km\ s}^{-1}$, and an abundance cut-off at $1.25\times10^{15}$\,cm, analagous to our assumptions for PN, we obtain a tentative abundance $X(\mathrm{PO}/\mathrm{H_2})$ in the range $5.4\times10^{-8} - 6.0\times10^{-7}$.

As for PN, we find that $T_{\mathrm{rot}}<<T_{\mathrm{kin}}$ and hence, that PO is subthermally excited. Also here, a detailed radiative transfer analysis would be needed to say more about this.

\subsection{Comparison to VY~CMa}
\cite{tenenbaum2007}  reported an abundance  $X(\mathrm{PO}/\mathrm{H}_2)\approx9\times10^{-8}$ for VY~CMa, comparable to what they found for PN. In a new study of PO, combining the single-dish observations of \cite{tenenbaum2007} and the SMA survey data of VY~CMa,  we find a column density of $\sim$$2\times10^{16}$\,cm$^{-2}$, corresponding to $X(\mathrm{PO/H_2})\approx7\times10^{-8}$. The size of the emitting region could be constrained to a radial extent of $\sim$$32\,R_{\star}$. For both IK~Tau and VY~CMa, the derived abundances and radial extents of the PO and PN-bearing gas are comparable. Observations at higher spatial resolution will, however, be needed to put stronger constraints on this, as the emission from both molecules is not spatially resolved in the observations with the SMA.

\begin{table} 
\caption{
PO components within the observed frequency ranges, ordered according to increasing frequency.
}\label{tbl:PO_components}
\centering
\small
\begin{tabular}{cccccccc}
\hline\hline\\[-2ex]
$\nu$&$F^{\prime}$&$F$ & $\Lambda$& $E_{\mathrm{up}}/k$&$A_{u\rightarrow l}$ &  $N_{\mathrm{HFS}}$\\
(GHz)	& 	&	&  &(K)	&(s$^{-1}$)	 &\\
\hline\\[-2ex]
\multicolumn{3}{l}{$^{2}\Pi_{1/2}\, J=5/2-3/2$}&& 8.4&&\\
108.707192 &    2	&2	&$f$&&  $2.13\times10^{-6}$	&1\\
108.998445 &    3	&2 	&$e$&&  $2.13\times10^{-5}$	&2\\
109.045396 &    2	&1 	&$e$&&  $1.92\times10^{-5}$	&3\\
109.206200 &    3	&2 	&$f$&&  $2.14\times10^{-5}$ &4\\
109.271376 &    2	&2 	&$e$&& $ 2.14\times10^{-6}$	&5\\
109.281189 &    2	&1 	&$f$&& $ 1.93\times10^{-5}$	&6\\

\multicolumn{3}{l}{$^{2}\Pi_{3/2}\,  J=5/2-3/2$}&& 328.7&&\\
109.588936 &2	& 2	&$f$&&  $1.45\times10^{-6}$ 	&1\\
109.589022 &2	& 2	&$e$&&  $1.45\times10^{-6} $ 	&2\\
109.829923 &3	& 2	&$f$&&  $1.45\times10^{-5} $	&3\\
109.829923 &3	& 2	&$e$&&  $1.45\times10^{-5} $ 	&4\\
109.966036 &2	& 1	&$f$&&  $1.31\times10^{-5} $	&5\\
109.966036 &2	& 1	&$e$&&  $1.31\times10^{-5} $ 	&6\\

\multicolumn{3}{l}{$^{2}\Pi_{1/2}\,  J=7/2-5/2$}&& 15.8&&\\
152.389058 &3	&3 &$f$&&  $2.99\times10^{-6} $	&1\\
152.656979 &4	&3 &$e$&&  $6.27\times10^{-5}$ 	&2\\
152.680282 &3	&2 &$e$&&  $5.98\times10^{-5}$ 	&3\\
152.855454 &4	&3 &$f$&&  $6.30\times10^{-5}$ 	&4\\
152.888128 &3	&2 &$f$&&  $6.00\times10^{-5}$ 	&5\\
152.953247 &3	&3 &$e$&&  $2.99\times10^{-6}$ 	&6\\

\multicolumn{3}{l}{$^{2}\Pi_{3/2}\,  J=7/2-5/2$}&& 336.0&&\\
153.616158 &3	&3 &$f$&&  $2.54\times10^{-6}$ 	&7\\
153.616330 &3	&3 &$e$&&  $2.54\times10^{-6}$ 	&8\\
153.793520 &4	&3 &$f$&&  $5.35\times10^{-5}$ 	&9\\
153.793520 &4	&3 &$e$&&  $5.35\times10^{-5}$ 	&10\\
153.857155 &3	&2 &$f$&&  $5.09\times10^{-5}$ 	&11\\
153.857155 &3	&2 &$e$&&  $5.09\times10^{-5}$ 	&12\\

\multicolumn{3}{l}{$^{2}\Pi_{1/2}\,  J=13/2-11/2$}&& 50.3&&\\
 283.3486949  & 6  & 6 &$f$& 	&   $5.55\times10^{-6}$ &1\\
 283.5868346  & 7  & 6 &$e$&    	&   $4.33\times10^{-4}$ &2\\
 283.5931860  & 6  & 5 &$e$&    	&   $4.27\times10^{-4}$ &3\\
 283.7776104  & 7  & 6 &$f$&    	&   $4.34\times10^{-4}$ &4\\
 283.7854186  & 6  & 5 &$f$&    	&   $4.28\times10^{-4}$ &5\\
 283.9124850  & 6  & 6 &$e$&    	&  $ 5.56\times10^{-6}$ &6\\

\multicolumn{3}{l}{$^{2}\Pi_{3/2}\,  J=13/2-11/2$}&& 370.8&&\\
 285.5285053  & 6  & 6 &$f$&   	&   $5.40\times10^{-6}$ &1\\
 285.5291059  & 6  & 6 &$e$&   	&   $5.40\times10^{-6}$ &2\\
 285.6260820  & 7  & 6 &$f$&   	&   $4.21\times10^{-4}$ &3\\
 285.6260820  & 7  & 6 &$e$&   	&   $4.21\times10^{-4}$ &4\\
 285.6438600  & 6  & 5 &$f$&   	&   $4.16\times10^{-4}$ &5\\
 285.6438600  & 6  & 5 &$e$&   	&   $4.16\times10^{-4}$ &6\\

\multicolumn{3}{l}{$^{2}\Pi_{1/2}\, J=15/2-13/2$}&& 66.0&&\\
 326.9859810  & 7  & 7 &$f$&    	&   $6.41\times10^{-6}$ &1\\
 327.2193520  & 8  & 7 &$e$&    	&   $6.73\times10^{-4}$ &2\\
 327.2239530  & 7  & 6 &$e$&    	&   $6.66\times10^{-4}$ &3\\
 327.4093210  & 8  & 7 &$f$&    	&   $6.74\times10^{-4}$ &4\\
 327.4149140  & 7  & 6 &$f$&    	&   $6.67\times10^{-4}$ &5\\
 327.5496150  & 7  & 7 &$e$&    	&  $ 6.41\times10^{-6}$ &6\\

\multicolumn{3}{l}{$^{2}\Pi_{3/2}\,  J=15/2-13/2$}&& 386.7&&\\
 329.4735908  & 7  & 7 &$f$&  	&   $6.31\times10^{-6}$ &7\\
 329.4743912  & 7  & 7 &$e$&   	&   $6.31\times10^{-6}$ &8\\
 329.5578520  & 8  & 7 &$f$&   	&   $6.63\times10^{-4}$ &9\\
 329.5578520  & 8  & 7 &$e$&   	&   $6.63\times10^{-4}$ &10\\
 329.5712820  & 7  & 6 &$f$&   	&   $6.56\times10^{-4}$ &11\\
 329.5712820  & 7  & 6 &$e$&   	& $  6.56\times10^{-4}$ &12\\
 \hline
\end{tabular}
\normalsize
\tablefoot{The respective columns list the rest frequency $\nu$ (obtained from CDMS), $F^{\prime}$ and $F$ of the upper and lower levels, respectively, the upper-level energy $E_{\mathrm{up}}/k$, the name of the $\Lambda$ component, the Einstein-$A$ coefficients, and $N_{\mathrm{HFS}}$, the HFS component number used for labelling purposes in Figs.~\ref{fig:PO_components_SMA} and \ref{fig:PO_components_IRAM}.}
\end{table}

\begin{table} 
\caption{Overview of the observed PO $^2\Pi_{\Omega}$ transitions.}\label{tbl:PO_rotdiagram}
\centering
\begin{tabular}{cccrcc}\hline\hline\\[-2ex]
				&			&				 	&												 & Rms		&  \\
$\Omega$	&$J^{\prime}-J$	&	$\Lambda$	&	 \multicolumn{1}{c}{Line flux}	&noise			& $\delta\nu$\\
				&						&								&\multicolumn{1}{c}{(K\,km\,s$^{-1}$)}		&(mK)			& (MHz)\\
\hline\\[-2ex]
$1/2$&$5/2-3/2$			& $e$						&	29.191								&	1.7		&2.0\\
$1/2$&$5/2-3/2$			& $f$						&	32.046								&	1.7		&2.0\\
$3/2$&$5/2-3/2$			& $e$						&	0.030								&	2.1		&2.0\\
$3/2$&$5/2-3/2$			& $f$						&	0.030								&	2.1		&2.0\\
$1/2$&$7/2-5/2$			& $e$						&	33.269								&	1.7		&2.0\\
$1/2$&$7/2-5/2$			& $f$						&	27.782								&	1.7		&2.0\\
$3/2$&$7/2-5/2$			& $e$						&	1.622								&	2.0		&2.0\\
$3/2$&$7/2-5/2$			& $f$						&	1.622								&	2.0		&2.0\\
\hline\\[-2ex]
			&				&								&(Jy\,km\,s$^{-1}$)							& (mJy)	& 	\\
\hline\\[-2ex]
$1/2$&$13/2-11/2$			& $e$						&1.528									&	33.4	&2.4\\
$1/2$&$13/2-11/2$			& $f$						&	1.183								&	33.4	&2.4\\
$3/2$&$13/2-11/2$			& $e$						&	0.471								&	27.8	&2.4\\
$3/2$&$13/2-11/2$			& $f$						&	0.471							&	27.8	&2.4\\
$1/2$&$15/2-13/2$			& $e$						&	0.657								&	48.8	&2.4\\
$1/2$&$15/2-13/2$			& $f$\tablefootmark{($\dagger$)}	&0.133				&	48.8	&2.4\\
$3/2$&$15/2-13/2$			& $e$						&	0.196							&	50.6	&2.4\\
$3/2$&$15/2-13/2$			& $f$						&	0.196							&	50.6	&2.4\\
\hline
\end{tabular}
\tablefoot{The respective columns list the quantum number $\Omega$, the rotational transition $J^{\prime}-J$, the $\Lambda$ component, the velocity-integrated line flux, the rms noise, and the frequency resolution $\delta\nu$.\\
\tablefoottext{$\dagger$}{Component 1 (see Table~\ref{tbl:PO_components}) was not fitted in our analysis, because the noise was much higher owing to an optically thick line of terrestrial water at $\sim$325\,GHz.  Instead, component 1 was assumed to have the same strength as component 6. }
}
\end{table}

\section{Discussion}\label{sect:discussion}
\subsection{Current modelling status on phosphorous chemistry}
From the LTE calculations of \cite{tsuji1973} it follows that PS should be the major carrier of P in the gas phase of an O-rich CSE, and that PO, PN, and PH are of much less importance.  In case of very high pressures PH$_3$ would be the major P-sink in the gas phase\footnote{Further details on or explanation of this were not given by \cite{tsuji1973}.}.

\cite{willacy1997} presented a model for the chemistry in the CSE of O-rich AGB stars at radii larger than $R_0=2\times10^{15}$\,cm, corresponding to about $60\,R_{\star}$ in the case of IK~Tau. At the inner radius of their model, they injected PH$_3$ as a parent molecule, with an abundance of $3\times10^{-8}$. Atomic P, P$^+$, PH, and PH$_2$ are the most abundant phosphorous species in their model, while PO reaches an abundance peak of $\sim$$3\times10^{-10}$, and only has an abundance above $10^{-10}$ for a narrow shell in the range $1-3\times10^{16}$\,cm, i.e. $700-2100\,R_{\star}$ in the case of IK~Tau.

\cite{mackay2001_PinCSEs} described the chemistry at radii larger than $R_0=1\times10^{15}$\,cm, corresponding to about $30\,R_{\star}$ in the case of IK~Tau. The chemical processes in this region of the envelope are strongly linked to photodissociation of the CSE material. They assumed PS to be the dominant P-bearing species at $R_0$ in the case of O-rich stars. The abundance of PS injected into the outer layers of the CSE ($r>R_0$) was assumed to correspond to 10\% of the cosmic abundance of P, where all the rest is locked up in solid dust. They found that only low abundances (peaking at $\sim$$10^{-10}$) of PO could be produced in the outer envelope and that essentially no other P-bearing species would be formed. They report a high atomic P abundance in the photodissociation region, which coincides with the presence of reactive species such as the OH radical. Neutral-neutral processes could in this case, if they are sufficiently fast, contribute to the formation of P-bearing molecules.

\cite{agundez2007_hcp} presented thermochemical equilibrium models for the region $1-9\,R_{\star}$ for O-rich and C-rich AGB stars (see their Fig.~2). In the O-rich model PO, PS and P$_2$ reach abundances up to $\sim$$10^{-7}$ at radii below $4\,R_{\star}$. The abundances of PN and PH$_3$ do not exceed $10^{-10}$ and $10^{-12}$, respectively. At radii $4\,R_{\star}<r<9\,R_{\star}$, P$_4$O$_6$ is the dominant carrier of phosphorus, with an abundance $\sim$$10^{-7}$, while all other P-bearing species become insignificant. Unfortunately, \cite{agundez2007_hcp} did not present modelling results regarding nonequilibrium chemistry at larger radii in O-rich envelopes.

\cite{cherchneff2012_irc10216revisited} modelled in detail the chemistry in the inner envelope of the C-rich IRC~+10\,216 as it is influenced by pulsational shocks passing through these layers. She found that PN can be formed very efficiently in the post-shock gas, reaching an abundance of $4.3\times10^{-7}$ at $5\,R_{\star}$. The formation path of PN is in this case dominated by reactions involving CN and CP. \cite{cherchneff2006} reported that the abundance of CN in the post-shock gas (at $1\,R_{\star}$) is comparable in O-rich and C-rich stars. Phosphorous species were, however, not included in this older model. It is hence difficult to assess the potential efficiency of PN formation in the inner wind of a shocked O-rich environment. \cite{cherchneff2012_irc10216revisited} did not include PO, PS, or PH$_3$ in her recent study of IRC~+10\,216's inner wind, all molecules that might be potentially important in the O-rich case.

From the above mentioned models, it is clear that PH$_3$, P$_4$O$_6$, PS, and PO are considered the main sinks for gas phase phosphorus. Predicted fractional abundances (with respect to H$_2$) are limited to $\sim$$10^{-9}$, leaving all species but PO barely detectable\footnote{We do not consider P$_4$O$_6$ in the further discussion. It has a tetrahedral symmetry, and due to the lack of a permanent dipole moment pure rotational transitions in the (sub)millimeter domain are not observable \citep{carbonniere2008}.}. The main uncertainties for the P-network in the current chemical models are the depletion factor of P, and the rate coefficients and activation energies for many of the involved reactions. Often, the isovalence of P and N is assumed to derive these rates and energies, but the validity of this assumption has been questioned \citep{cherchneff2012_irc10216revisited}.

\subsection{Comparison to observations}
In contrast to the model calculations, we established that PN and PO are about equally abundant in the winds of the O-rich supergiant VY~CMa and of the O-rich AGB star IK~Tau. The emission of both species arises from a region with a maximum radial extent of $32-41\,R_{\star}$ in the case of VY~CMa and an estimated $40\,R_{\star}$ in the case of IK~Tau. For VY~CMa the high abundances of PO and PN, with respect to the available chemical models, were tentatively explained by the fact that the CSE might be more active than that of the average O-rich AGB star assumed in the chemical models. However, as pointed out by \cite{milam2008_phosphorus} and confirmed by the SMA observations, PO and PN are both present in the more quiescent spherical part of VY~CMa's CSE, which is quite probably comparable to the CSE of an O-rich AGB star. Since IK~Tau is believed to be a rather well behaved Mira-type O-rich AGB star, and we can now firmly establish the presence of PO and PN with much higher than predicted abundances in its CSE, it is clear that there is a strong need for an updated treatment of the chemical networks involving P-bearing species.

We estimate that PO and PN are produced at radii $r\lesssim40\,R_{\star}$ in the CSEs of both IK~Tau and VY~CMa. Very close to the star, shock chemistry \citep[e.g.][]{cherchneff2006,cherchneff2012_irc10216revisited} could play an important role. Further out, where dust is forming or has formed, dust-gas chemistry could become important. To date, however, this type of chemical modelling has not been presented for AGB stars, owing to the many uncertainties involved in the interpretation of dust-gas interaction. Photodissociation chemistry typically becomes important at radii $r\approx10^{16}\,\mathrm{cm}$ (a few 100 stellar radii). The latter is thus, in first instance, not as urgent to understand in terms of phosphorous chemistry as the first two. It is, however, clear from the literature and from this work, that the chemical network involving phosphorus is still very uncertain and that further efforts are needed to understand the phosphorous chemistry throughout the circumstellar environments of evolved stars of all chemical types.


\section{Conclusions}\label{sect:conclusions}
We present the first observations of PN and PO emission towards an oxygen-rich AGB star. We analysed dedicated line observations with the IRAM~30\,m telescope and data obtained in the course of an unbiased SMA survey which covers the range $279-355$\,GHz. Emission is detected in four rotational transitions for each species: $J= 2-1,3-2,6-5$, and $7-6$ for PN, and $J=5/2-3/2,7/2-5/2,13/2-11/2$, and $15/2-13/2$ for PO. We detected 13 different spectral features corresponding to different fine and hyperfine components of the PO transitions. From the observed lines we estimate an abundance $X(\mathrm{PN/H_2})=2.8\pm1.5\times10^{-7}$ and we derive $X(\mathrm{PO/H_2})$ to lie in the range $5.4\times10^{-8} - 6.0\times10^{-7}$. A detailed radiative transfer study is needed to further constrain these abundances, but in order to successfully do this, observations with higher S/N are needed, as well as collision rate coefficients. Observations at higher spatial resolution will also help constrain the spatial distribution of both molecules in the CSE. Full science capabilities of the Atacama Large (sub)Millimetre Array (ALMA) will lead to a spatial resolution in the range $0\farcs006-0\farcs037$, corresponding to about $0.7-5\,R_{\star}$ for IK~Tau and VY~CMa. We should hence be able to trace where the molecules are produced and destroyed, directly constraining the potential chemical formation/destruction routes.

Our abundance estimates for PO and PN exceed by several orders of magnitude the abundances predicted by models of phosphorous chemistry in O-rich CSEs. At the same time, no lines of PS or PH$_3$ are detected, although these molecules are often predicted or assumed to be the more abundant P-bearing species in these CSEs, and the parent molecules in the P-network. We suggest that PO and PN are the major gas-phase reservoirs of phosphorus in O-rich environments like IK~Tau's and VY~CMa's CSEs. 

The influence of shocks on the inner-wind chemistry of O-rich evolved stars should be further studied and extended to include chemical networks involving P-bearing species. Reaction rates and activation energies of chemical reactions involving P-bearing species should be (better) determined. Another large uncertainty in the current models that treat P-bearing species is the depletion factor of P throughout the CSE. 

It will be important to attempt to trace the parent molecules in the P-network. For O-rich stars there should be a focus on further observing PO and PN, but also on PS and PH$_3$, since the latter two are often put forward as major P-carriers. For C-rich stars, there should most likely be a focus on HCP, CP, PH$_3$, and PN. It would hence be important to try and trace P-bearing species in the CSEs of a substantial sample of evolved stars, covering the different chemical types (O, S, C), but also the different pulsational types, in order to probe the influence of pulsation/shock strength on the phosphorous chemistry.


\begin{acknowledgements}
The Submillimeter Array is a joint project between the Smithsonian Astrophysical Observatory and the Academia Sinica Institute of Astronomy and Astrophysics and is funded by the Smithsonian Institution and the Academia Sinica.
\end{acknowledgements}


\addcontentsline{toc}{chapter}{Bibliography}
\bibliographystyle{aa}
\bibliography{DeBeck_21349.bib}


\end{document}